\documentclass[3p]{elsarticle}
\usepackage{epstopdf}
\usepackage{subfigure,subcaption,graphicx,float}
\usepackage{amsmath, amsfonts, amssymb}
\usepackage{amsthm}
\usepackage{yhmath}
\usepackage{epsf}
\usepackage{amsfonts}
\usepackage{stmaryrd}
\usepackage{amssymb}
\usepackage{leftidx}
\usepackage{xcolor}
\usepackage{mathtools}
\usepackage{placeins}
\usepackage{booktabs}
\usepackage{enumitem}
\usepackage{caption}
\usepackage{tabu}
\usepackage{multirow}
\usepackage{stackengine}
\usepackage[utf8]{inputenc}
\usepackage[english]{babel}
\usepackage{bm}
\usepackage{array}
\usepackage{multirow}

\def\bfy{{\bf y}}

\def\bfN{{\bf N}}

\def\bfS{{\bf S}}

\def\bfX{{\bf X}}
\def\bfF{{\bf F}}

\def\eps{\varepsilon}

\def\e0{\varepsilon_0}

\def\s0{\sigma_0}

\DeclareMathAlphabet{\mathsfit}{T1}{\sfdefault}{\mddefault}{\sldefault}
\SetMathAlphabet{\mathsfit}{bold}{T1}{\sfdefault}{\bfdefault}{\sldefault}

\theoremstyle{plain}

\long\def\symbolfootnote[#1]#2{\begingroup%
\def\thefootnote{\fnsymbol{footnote}}\footnote[#1]{#2}\endgroup}

\begin{document}
\begin{frontmatter}

\title{On failure mechanisms and load-parallel cracking in confined elastomeric layers \vspace{0.1cm}}

\vspace{-0.1cm}

\author{Aarosh Dahal}
\ead{adahal8@gatech.edu}

\author{Aditya Kumar\corref{cor1}}
\ead{aditya.kumar@ce.gatech.edu}

\address{School of Civil and Environmental Engineering, Georgia Institute of Technology, Atlanta, GA 30332, USA \vspace{0.05cm}}

\cortext[cor1]{Corresponding author}

\begin{abstract}

\vspace{-0.1cm}

Thin layers of elastomers bonded to two rigid plates demonstrate unusual failure response. 
Historically, it has been believed that strongly-bonded layers fail by two distinct mechanisms: (i) internal/external penny-shaped crack nucleation and propagation, and (ii) cavitation, that is, cavity growth leading to fibrillation and then failure. However, recent work has demonstrated that cavitation itself is predominantly a fracture process. 
While the equations describing cavitation from a macroscopic or top-down view are now known and validated with experiments, several aspects of the cavitation crack growth need to be better understood. 
Notably, cavitation often involves through-thickness crack growth parallel to the loading direction, raising questions about when it initiates instead of the more typical penny-shaped cracks perpendicular to the load.
Understanding and controlling the two vertical and horizontal crack growth is key to developing tougher soft films and adhesives.
The purpose of this Letter is to provide an explanation for the load-parallel crack growth through a comprehensive numerical analysis and highlight the role of various material and geometrical parameters.

\endkeyword

\end{abstract}

\end{frontmatter}

\section{Introduction}

Thin layers of soft solids are used in various applications, such as pressure-sensitive adhesives, protective coatings, barrier packaging films, and flexible electronics. In many of these applications, they are often constrained on both sides.  The mechanical response of constrained thin layers has been widely investigated in the literature, both experimentally and theoretically. 
The study of failure has especially proven to be challenging because of the associated nonlinearity and the sensitive dependence on geometrical and material parameters.

\begin{figure}[t!]
	\centering
	\includegraphics[width=6.3in]{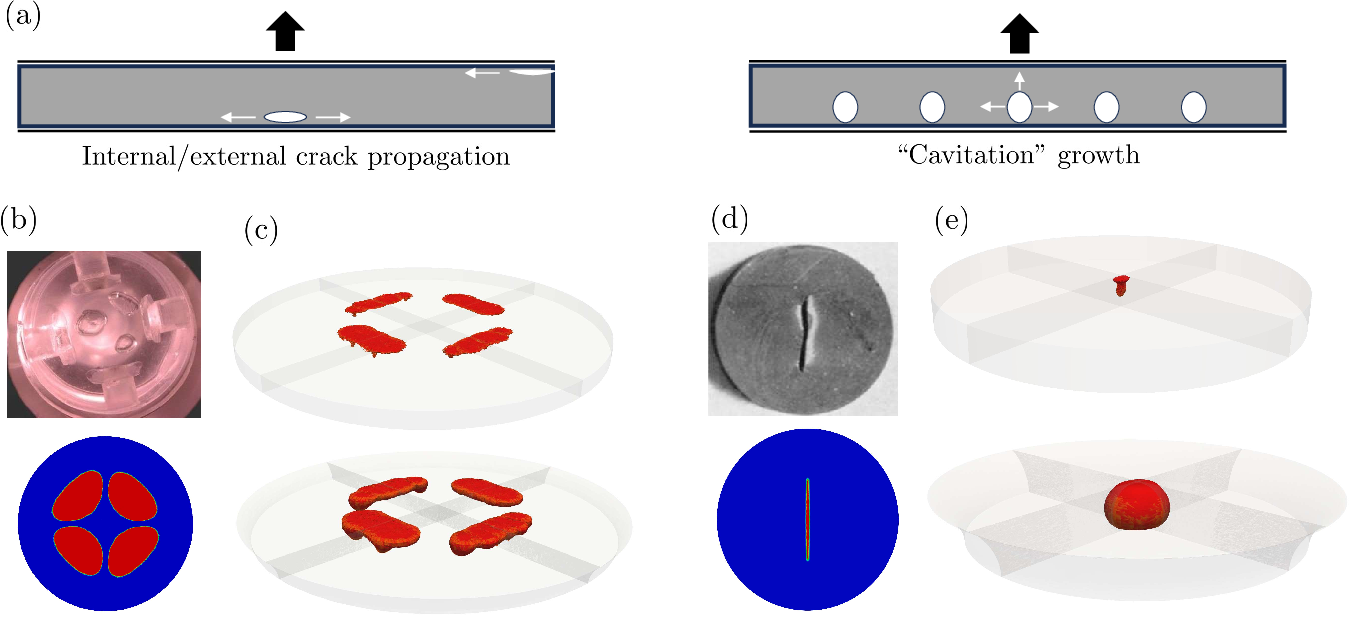}
	\caption{(a) Two mechanisms for failure of thin bonded layers: internal/external penny-shaped crack propagation and ``cavitation" growth. (b) Results from the comparison presented in \cite{KKLP24} for thin poker-chip specimens with the experimental results of \cite{guo-ravichandar23} demonstrating penny-shaped crack growth. (c) Results with the phase-field theory presented in Section 2 for a silicone elastomer demonstrating penny-shaped crack growth in undeformed (top) and deformed (bottom) configurations. (d) Results from the comparison presented in \cite{KLP21} for thin poker-chip specimens with the experimental results of \cite{GentLindley1959} demonstrating cavitation crack growth. (e) Results with the phase-field theory presented in Section 2 for natural rubber showing cavitation crack growth in undeformed (top) and deformed (bottom) configurations.}\label{Fig1}
\end{figure}

In thin layers bonded and constrained between stiff fixtures and subjected to tension, cracks are typically observed to nucleate at or near the interface where the hydrostatic stress is the highest. Based on extensive experimental testing \cite{GentLindley1959, creton1999direct, creton2000flatprobe, crosbyshullcreton2000}, two distinct failure mechanisms have been identified: (i) nucleation of an internal penny-shaped crack in the center or on the lateral surface or the nucleation of an adhesive crack that propagates perpendicular to the direction of loading, (ii) nucleation of a crack/cavity that grows both parallel and perpendicular to the loading direction. The mechanisms are illustrated in Fig.~\ref{Fig1}. Crosby et al. \cite{crosbyshullcreton2000} discussed how the two mechanisms may depend on the various material and geometrical parameters. 
They and others analyzed the first mechanism using Griffith's fracture theory \cite{Griffith21}. On the other hand, the analysis of the second mechanism, known as cavitation, was made using elastic instability theory \cite{GentLindley1959}. The prevailing notion has been that cavitation growth initiates as an elastic instability, causing the formation of a cavity that expands into the bulk of the material parallel to the loading direction. 

%
However, over the last decade, through direct comparisons with a comprehensive set of cavitation experiments, it has been established that cavitation is, first and foremost, a \emph{fracture} process \cite{lefevre2015cavitation, Poulain17, KFLP18, KLP21, KKLP24}. For a detailed review of the evidence against the elasticity theory of cavitation, the reader is referred to Breedlove et al. \cite{breedlove2024}.
To model cavitation nucleation and growth as a fracture phenomenon, Kumar et al. \cite{KFLP18} put forth a unified theory of fracture -- regularized, of phase-field type, that can model not only cavitation but also the other experimental results accumulated in the literature in the last 100 years on the nucleation and propagation of cracks within the bulk of elastic brittle elastomers as well as from pre-existing cracks, notches and other sources of stress concentrations.  
In this theory, the crack nucleation under uniform stresses is controlled only by the material's strength surface \cite{KBFLP20}. 
Notably, in this theory, the hydrostatic tensile strength of the material is considered an independent property, no different in its macroscopic form than the material strength under any other stress state. 
The growth of a large crack follows Griffith's theory. 

Using this theory, cavitation can be understood as a two-stage process. In the first stage, crack nucleation in very thin layers occurs according to the strength of the material. Interestingly, analysis of cavitation experiments in synthetic silicone rubber \cite{guo-ravichandar23} with this theory showed that the nucleation of cracks in thin layers is controlled not solely by the hydrostatic strength, but instead, the entire first octant (all principal stresses greater than zero) of the strength surface. 
In the second stage, the nucleated crack grows either parallel to the loading direction or perpendicular to it (Fig.~\ref{Fig1}(b)-(e)). In this stage, the propagating crack competes with the nucleation of new cracks that may lead to its arrest until sufficient energy is available for it to propagate again \cite{KLP21, ruihuang23cavitation, KKLP24}. The nucleation of new cracks is no longer governed solely by material strength, as non-uniform stress in the thin layer brings toughness into play.
With this theory, complete quantitative comparisons were conducted with the \emph{poker-chip} cavitation experiments in natural rubber and synthetic silicone \cite{KLP21, KKLP24}, which have shown very good agreement. 

With cavitation viewed as a fracture process, both internal/external crack growth and cavitation growth can be viewed through the same lens. They both involve crack nucleation and propagation and are controlled by the material's strength as well as its critical energy release rate. Instead, the central unanswered question is why, during cavitation growth in thin layers, cracks propagate vertically---parallel to the applied tensile load---rather than horizontally, as is more typical. This letter aims to explain that behavior. 
This is achieved through a comprehensive fracture analysis combining the phase-field model, which captures both stages of cavitation, and an energy release rate approach for analyzing the growth of nucleated cracks.
The analysis is presented in Section 3, following a brief discussion of the theory in the next section.

\section{The macroscopic theory for fracture nucleation and propagation}\label{Sec: Fracture}

Based on decades of experimental observations, the nucleation of fracture in nominally elastic brittle materials, whether soft or hard, are of three types \cite{KBFLP20}: when the stress state is spatially uniform, when a large pre-existing crack is present, and when neither holds true. Under uniform stress states, when no large defects are present in the structure, fracture nucleates when the stress reaches a strength threshold. The strength threshold in three dimensions is defined through a scalar function of the invariants of the stress tensor, called the strength surface, 
\begin{equation}
\mathcal{F}(\bfS)=0 ,\label{Strength surface}
\end{equation}
where $\bfS$ is a suitable measure of stress. The strength surface can be characterized from several multiaxial experiments for any brittle material. Some common forms of the strength surface include the Drucker-Prager surface and Mohr-Coulomb surface.

When the structure contains a large pre-existing crack, fracture may nucleate from the crack front and it may propagate. The nucleation of the fracture is defined through the Griffith criterion
\begin{equation}
-\dfrac{\partial{\mathcal{W}}}{\partial \Gamma} \leq G_c \label{Griffith},
\end{equation}
which captures the energy competition between the bulk elastic energy and the surface fracture energy. The surface fracture energy, according to Griffith \cite{Griffith21}, is proportional to the surface area created with $G_c$, the fracture toughness or the critical energy release rate, as the proportionality constant. 
The Griffith criterion, however, contains no direct information about crack propagation. It is essentially a stability criterion for the current position of the crack. 

The Griffith criterion was generalized by Francfort and Marigo \cite{Francfort98} and cast in a mathematically consistent variational formulation to also capture crack propagation. In their formulation, the deformation field $\bfy(\bfX,t)$ and the crack surface $\Gamma (t)$ at any given discrete time $t_k\in\{0=t_0,t_1,...,t_m,t_{m+1},...,t_M=T\}$ are obtained through minimizing the sum of elastic energy and fracture energy:
\begin{equation}
(\bfy_k,\Gamma_k)=\underset{\begin{subarray}{c}
  \bfy=\overline{\bfy}(t_k)\,{\rm on}\,\partial \Omega_0^D \\[1mm]
  \Gamma \supset \Gamma_{k-1}
  \end{subarray}}{\arg\min}\, \mathcal{E}(\mathbf{y}, \Gamma) := \int_{\Omega_0 \setminus \Gamma} W(\mathbf{F}) \, \mathrm{d}\mathbf{X} 
+ G_c \mathcal{H}^{N-1}(\Gamma) 
\label{Variational} 
\end{equation}
In this expression, $\mathcal{H}^{N-1}(\Gamma)$ stands for the ($N-1$)–dimensional Hausdorff measure (the surface measure) of the unknown crack where $N$ is the space dimension. $W(\mathbf{F})$ stands for the hyperelastic energy function of the deformation gradient tensor, $\bfF$. In this work, we typically adopt the compressible Neo-Hookean function
\begin{equation}
W(\bfF)=\mathcal{W}(I_1) + \kappa g(J)=\dfrac{\mu}{2}\left[I_1-3\right]-\mu\ln J+\dfrac{\kappa}{2}(J-1)^2,\label{W-NH}
\end{equation}
where
\begin{equation*}
I_1=\bfF\cdot\bfF \qquad {\rm and}\qquad J=\det\bfF
\end{equation*}
are the standard invariants, while $\mu$ and $\kappa$ are material constants. Through linearization, the Poisson's ratio can be related to these constants as $\nu=\kappa/(2\, (\mu+\kappa)$. The non-Gaussian Ogden hyperelastic function \cite{Ogden72} is also used below for a study.

The variational criterion above can predict the arbitrary growth of large pre-existing cracks in a two or three-dimensional body without invoking any additional criterion. However, it requires a regularization to become amenable to numerical solutions. For this purpose, a phase field, $z = z (\bfX, t)$, is introduced, which takes values in the range [0, 1] over a phase boundary of infinitesimal width $\varepsilon$. Precisely, $z = 1$ identifies regions of the sound elastomer, whereas $z < 1$ identifies regions of the elastomer that have been fractured. The regularized variational formulation is obtained as
\begin{eqnarray}
(\bfy^\eps_k,z^\eps_k)=\underset{\begin{subarray}{c}
  \bfy=\overline{\bfy}(t_k)\,{\rm on}\,\partial\Omega_0^D \\[1mm]
  0 \leq z\leq z_{k-1}\leq 1
  \end{subarray}}{\arg\min}\,\mathcal{E}^{\eps}(\bfy,z):= & \displaystyle
  \int_{\Omega_0} \left[(z^2+\eta) \, \mathcal{W}(I_1) + (z^2+\eta_\kappa) \kappa g(J)\right]\,{\rm d}\bfX \nonumber\\
  & \displaystyle +\dfrac{3 G_c}{8}\int_{\Omega_0}\left(\dfrac{1-z}{\eps}+\eps\nabla z\cdot\nabla z\right)\,{\rm d}\bfX,\label{BFM00}
\end{eqnarray}
The regularization of the surface energy $\Gamma$-converges to its sharp form for $\varepsilon \rightarrow 0$.
The parameters $\eta$ and $\eta_\kappa$ in the regularized formulation stand for small positive numbers that aid the numerical tractability of the
vanishingly small stiffness of the regions of the elastomer that have undergone fracture \cite{KFLP18}. 

The third type of fracture nucleation is when neither a large crack is present nor there is uniform stress state. Analysis of experimental observations indicate that fracture nucleation in this scenario is controlled by an ``interpolation'' of the strength criterion (\ref{Strength surface}) and the energy release rate criterion (\ref{Variational}). Unifying the two criteria has been challenging historically due to their fundamentally different nature. 
They have been unified recently within the phase-field framework by Kumar et al. \cite{KFLP18, KBFLP20}. The resulting model can describe nucleation and propagation of fracture for arbitrary geometries and monotonic loadings. 

In the unified phase-field fracture formulation, the deformation field and the phase field at any given discrete time $t_k$ are determined by the system of coupled partial differential equations 
\begin{equation}
\left\{\begin{array}{ll}
{\rm Div}\left[(z_{k}^2+\eta)\dfrac{\partial \mathcal{W}}{\partial \bfF}(\nabla \bfy_{k}) + (z_{k}^2+\eta_\kappa) \kappa \, \dfrac{\partial g}{\partial \bfF}(\nabla \bfy_{k})\right]=\textbf{0},& \,\bfX\in\Omega_0\vspace{0.2cm}\\
\bfy_k(\bfX)=\overline{\bfy}(\bfX,t_k), & \; \bfX\in\partial\Omega_0^{\mathcal{D}}\vspace{0.2cm}\\
\left[(z_{k}^2+\eta)\dfrac{\partial \mathcal{W}}{\partial \bfF}(\nabla \bfy_{k}) + (z_{k}^2+\eta_\kappa) \kappa \, \dfrac{\partial g}{\partial \bfF}(\nabla \bfy_{k})\right]\bfN=\textbf{0},& \; \bfX\in\partial\Omega_0^{\mathcal{L}}
\end{array}\right. \label{BVP-y-theory}
\end{equation}
and
\begin{equation}
\left\{\begin{array}{ll}
\hspace{-0.15cm} \dfrac{3}{4} {\rm Div}\left[\varepsilon\, \delta^\varepsilon \, G_c \nabla z_k\right]=2 z_{k} W(\nabla\bfy_k)+c_\texttt{e}(\bfX,t_{k})-\dfrac{3}{8} \dfrac{ \delta^\varepsilon \, G_c}{\varepsilon},&\, \mbox{ if } z_{k}(\bfX)< z_{k-1}(\bfX),\quad \bfX\in \Omega_0
\vspace{0.2cm}\\
\hspace{-0.15cm}
\dfrac{3}{4} {\rm Div}\left[\varepsilon\, \delta^\varepsilon \, G_c \nabla z_k\right]\geq 2 z_{k} W(\nabla\bfy_k)+c_\texttt{e}(\bfX,t_{k})-\dfrac{3}{8}\dfrac{\delta^\varepsilon \,G_c}{\varepsilon},&\, \mbox{ if } z_{k}(\bfX)=1\; \mbox{ or }\; z_{k}(\bfX)= z_{k-1}(\bfX)>0, \quad \bfX\in \Omega_0\vspace{0.2cm}\\
\hspace{-0.15cm}
z_k(\bfX)=0,&\, \mbox{ if } z_{k-1}(\bfX)=0, \quad \bfX\in \Omega_0\vspace{0.2cm}\\
\hspace{-0.15cm}\nabla z_k\cdot\bfN=0,& \, \bfX\in \partial\Omega_0
\end{array}\right. \label{BVP-z-theory}
\end{equation}
These equations are constructed by first obtaining the Euler-Lagrange equations of the variational principle (\ref{BFM00}) and adding an additional term $c_\texttt{e}(\bfX,t)$ to the evolution equation for the phase field. The term $c_\texttt{e}(\bfX,t)$ is a driving force whose specific constitutive prescription, as discussed in \cite{KBFLP20}, depends on the particular form of strength surface, while $\delta^\varepsilon$  is
a non-negative coefficient whose specific constitutive
prescription depends in turn on the particular form of $c_\texttt{e}(\bfX,t)$. The inequalities in (\ref{BVP-z-theory}) stem from the bounded and irreversible nature of phase-field evolution; see \cite{KFLP18} for more details. The parameter $\varepsilon$ is chosen to be arbitrarily smaller than the smallest characteristic length scale in the problem. Specific form for $c_\texttt{e}(\bfX,t)$ for an elastomer whose strength surface is well described by the Drucker-Prager surface was presented recently in \cite{KKLP24}.


\section{Analysis of load-parallel crack growth} \label{Sec: Results}

\begin{figure}[h!]
	\centering
	\includegraphics[width=6.3in]{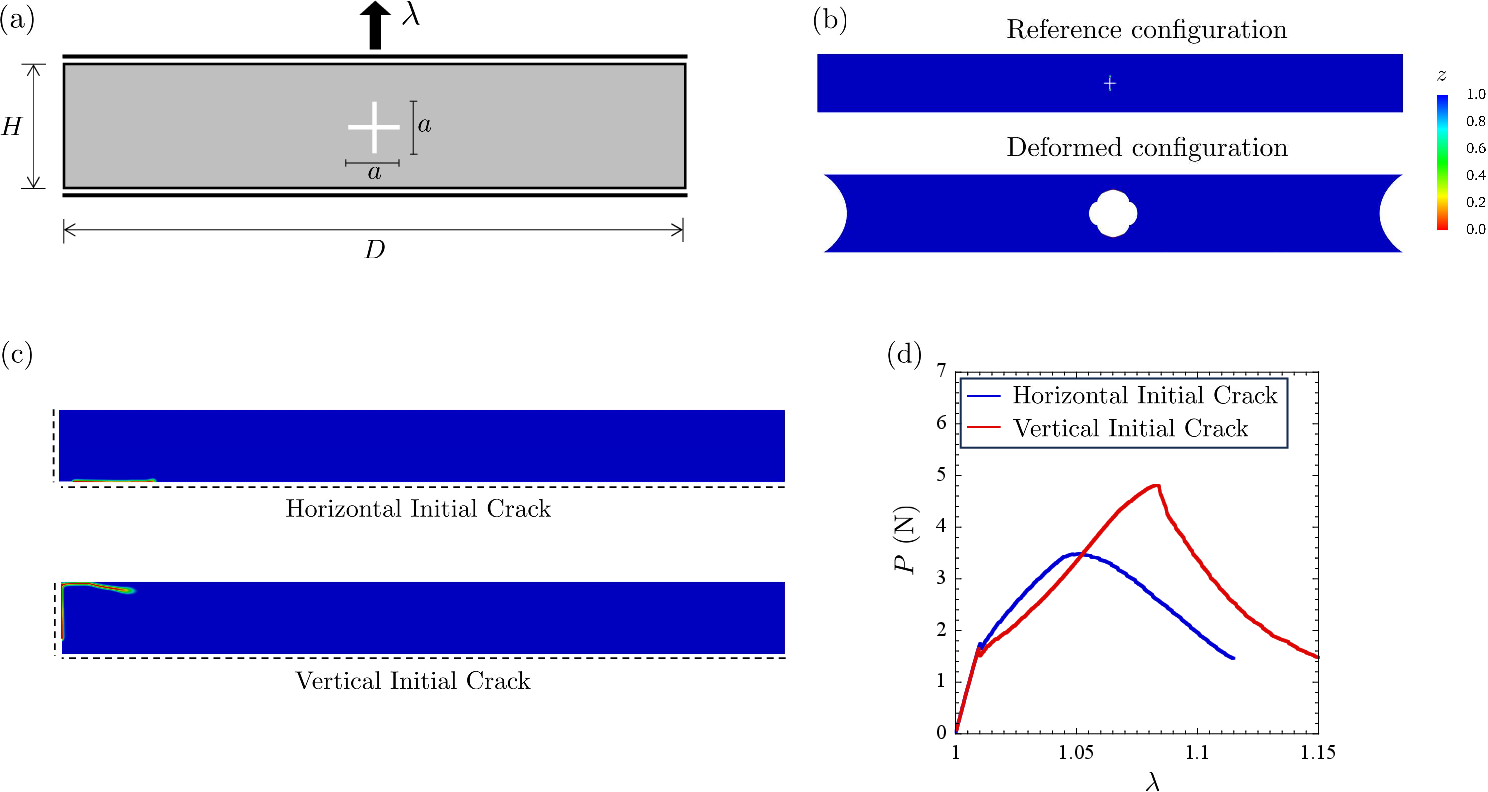}
	\caption{(a) Schematic of an idealized representation of a nucleated crack as a cross-shaped configuration in a thin bonded layer.  (b) Deformation and crack initiation from the cross-shaped crack configuration. (c) Crack initiation and propagation from horizontally and vertically aligned initial cracks. Phase-field contour plots are shown over one quarter of the geometry.  (d) Force-deformation response for the growth of horizontally and vertically aligned initial cracks.}\label{Fig2}
\end{figure}

As previously discussed, once a crack nucleates in a thin bonded layer, it can grow as a penny-shaped crack perpendicular to the loading or evolve into multiple crack fronts propagating in different directions. Fig.~\ref{Fig1}(c) and (e) show two phase-field simulation results, based on the theory from the previous section, that illustrate these distinct crack growth behaviors.  In the results shown in Fig.~\ref{Fig1}(e) for a thin layer made of natural rubber, the crack develops two fronts, one perpendicular and one parallel to the loading direction. Motivated by this result, we conduct a study in this section using the geometry shown in Fig.~\ref{Fig2}(a). We assume a single nucleation event at the center of a thin, two-dimensional layer, leading to a cross-shaped crack with both horizontal and vertical fronts. Our goal is to identify the conditions that favor vertical crack growth over horizontal growth. Under these conditions, a ``cavitation" growth would be observed experimentally, initially appearing as the formation of a nearly circular or spherical cavity in the deformed state; e.g., see Fig.~\ref{Fig2}(b).

The thin layer is modeled as a rectangular geometry in its undeformed, stress-free configuration, with a diameter $D = 10$ mm and a thickness $H$ in the range $H \in [0.5, 2]$ mm. We assume the interfaces to be infinitely strong, and consequently, the failure occurs internally within the bulk of the solid. A cross-shaped crack configuration with a crack length $a = 0.1$ mm is introduced at the center of the geometry. The length of each crack is taken to be larger than the intrinsic fracture length scale, such that their growth is governed solely by the material's critical energy release rate. This allows the calculation of the energy release rate for each crack over a wide range of deformations using the $J$-integral \cite{rice1968path}.
Note that the phase-field fracture model can predict the nucleation of cracks without any pre-seeding. However, for soft, thin films, multiple cracks may nucleate simultaneously, resulting in complex crack morphologies, which makes a systematic study of the entire parameter space and deformation regimes challenging. Hence, an idealized crack geometry is being chosen. Also note that cracks may, in general, develop more than two crack fronts, and crack kinking and fragmentation may also be observed \cite{guo-ravichandar23}.


%
\begin{figure}[h!]
	\centering
	\includegraphics[width=6.3in]{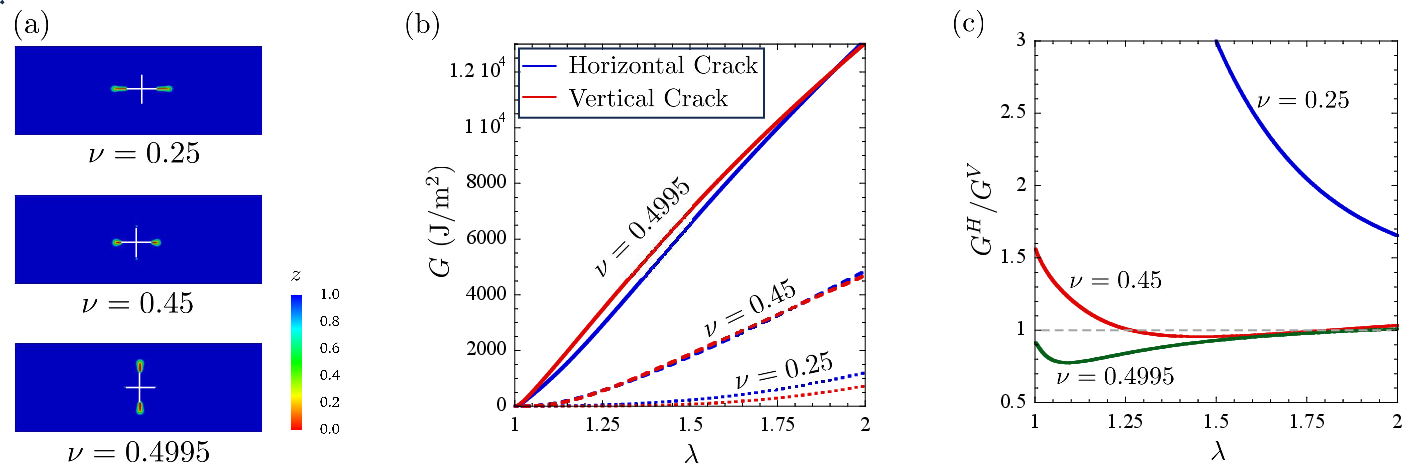}
	\caption{Influence of the material's Poisson's ratio, $\nu$, on the load-parallel crack propagation. (a) Contour plots of the phase field $z$ from simulations for $\nu= 0.25, 0.45$ and $4995$ and aspect ratio $D/H=10$. Only a cutout of the domain near the cracks is shown. (b) Energy release rate, $G$, as a function of applied stretch, $\lambda$, for the horizontal and vertical cracks, and (c) ratio of energy release rate of horizontal crack to vertical crack, $G^{H}/G^{V}$, as a function of $\lambda$, for the three values of $\nu$.}\label{Fig3}
\end{figure}

 To further highlight the impact of vertical versus horizontal crack growth on the layer's failure properties, phase-field simulation results are presented in Fig.~\ref{Fig2}(c)--(d) for two cases where either vertical or horizontal growth dominates. As observed from the load-deformation response, the vertical growth leads to a higher load-carrying capacity for the film, as indicated by the peak load in Fig.~\ref{Fig2}(d), and greater energy dissipation until failure, as represented by the area under the load-deformation curve.


\subsection{The effect of Poisson's ratio $\nu$}

We first study the influence of material compressibility on load-parallel crack propagation by varying the Poisson's ratio from highly compressible ($\nu=0.25$) to nearly incompressible ($\nu=0.4995$). The elastic behavior is taken to be described by the Neo-Hookean model (\ref{W-NH}).
Fig.~\ref{Fig3}(a) presents phase-field results from the simulation of a thin layer with an aspect ratio $D/H = 10$ and $G_c/\mu = 0.1$ mm for $\nu = 0.25$, $0.45$, and $0.4995$. The results show that for $\nu = 0.25$ and $0.45$, horizontal crack growth is preferred, while for $\nu = 0.4995$, the vertical crack growth occurs first.

Fig.~\ref{Fig3}(b)--(c) provide a more comprehensive analysis through the computation of the energy release rates for the horizontal crack, $G^H$, and the vertical crack, $G^V$. The energy release rates are plotted in Fig.~\ref{Fig3}(b) as a function of the applied stretch, $\lambda$. In Fig.~\ref{Fig3}(c), the ratio $G^H/G^V$ is plotted as a function of $\lambda$. A ratio $G^H/G^V < 1$ indicates a preference for vertical crack growth if crack nucleation occurs at the corresponding value of $\lambda$. If this ratio is close to 1, it indicates that both vertical and horizontal growth occur simultaneously.

We observe that for a highly compressible material, $G^H/G^V$ remains significantly greater than 1. On the other hand, for a nearly incompressible material, $G^H/G^V$ is notably less than 1 over a large range of $\lambda$ values.
From the $G$–$\lambda$ relationship shown in Fig.~\ref{Fig3}(b), one can deduce that for materials with a critical energy release rate $G_c < 5000$ J/m$^2$, the vertical crack growth is strongly preferred. For materials with intermediate compressibility ($\nu = 0.45$), horizontal crack propagation is favored at low $G_c$ values; otherwise, vertical and horizontal cracks may grow simultaneously.

\begin{figure}[h!]
	\centering
	\includegraphics[width=6.0in]{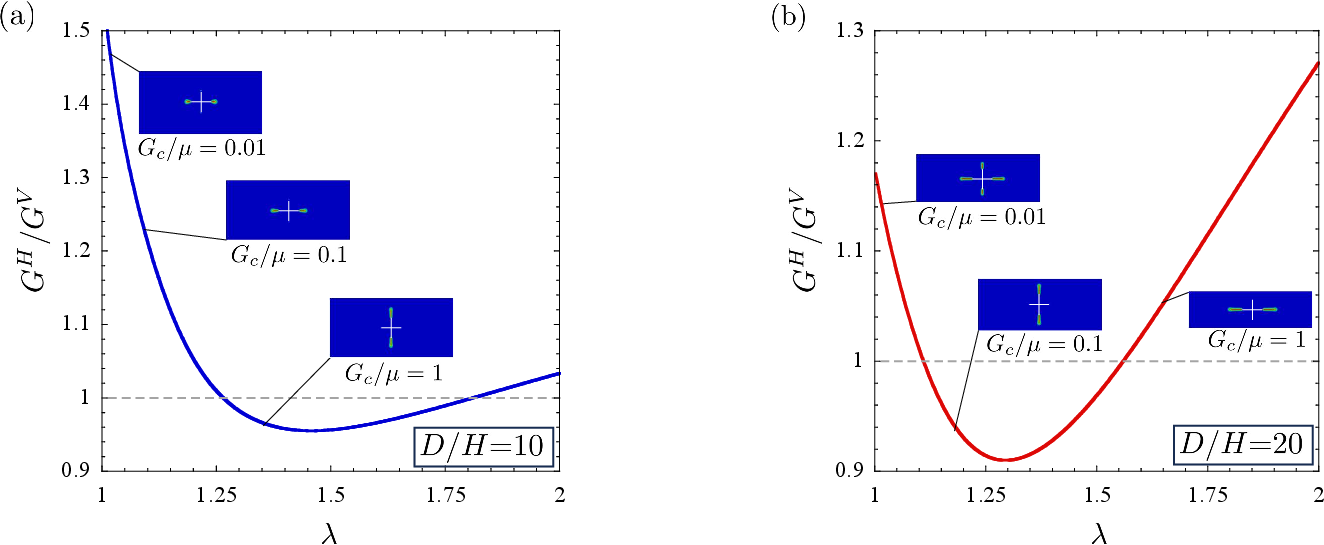}
	\caption{Influence of the ratio of fracture toughness to shear modulus, $G_c/\mu$, on load-parallel crack propagation. (a) Ratio of energy release rate of horizontal crack to vertical crack, $G^{H}/G^{V}$, as a function of $\lambda$ for the aspect ratio $D/H=10$ and Poisson's ratio $\nu=0.45$. Contours of the phase-field showing crack initiation from either horizontal or vertical cracks are included, corresponding to three values of $G_c/\mu= 0.01, 0.1$, and $1$ mm. (b)  $G^{H}/G^{V}$ as a function of $\lambda$ for  $D/H=20$. Contours of the phase field are shown corresponding to three values of $G_c/\mu= 0.01, 0.1$, and $1$ mm.)}\label{Fig5}
\end{figure}

To further demonstrate the effect of $G_c$, or more precisely, the ratio of $G_c$ to $\mu$, on load-parallel crack propagation, we conduct phase-field simulations for $G_c/\mu \in [0.01, 0.1, 1]$ mm. The results are presented in Fig.~\ref{Fig5} for $D/H = 10$ and $D/H = 20$ with $\nu = 0.45$, along with the computations for $G^H/G^V$ for a wide range of deformations. The two sets of results are consistent, showing that for $G^H/G^V < 1$, the vertical crack initiates first, whereas for $G^H/G^V > 1$, the horizontal crack initiates.

\subsection{The effect of geometrical parameters}

\begin{figure}[h!]
	\centering
	\includegraphics[width=6.5in]{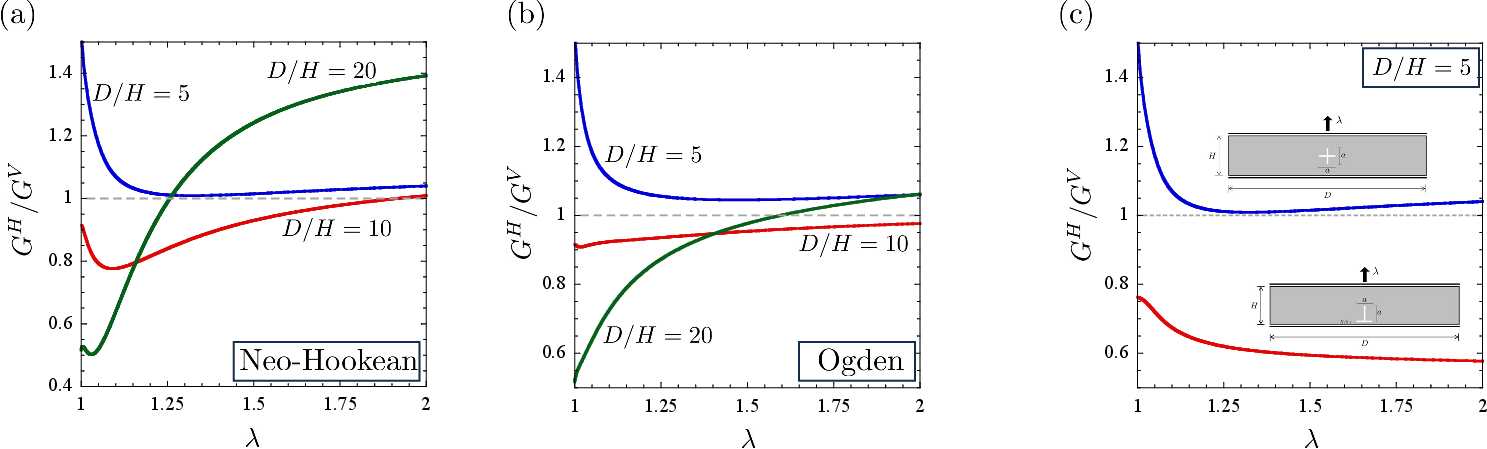}
	\caption{Influence of geometrical parameters on load-parallel crack propagation. (a) Ratio of energy release rate of horizontal crack to vertical crack, $G^{H}/G^{V}$, in a nearly incompressible Neo-Hookean material as a function of $\lambda$, for three values of the aspect ratio of the thin film, $D/H= 5, 10$ and $20$. (b) $G^{H}/ G^{V}$ as a function of $\lambda$ in an Ogden material with stiffening parameter $\alpha=4$. (c) Comparison of $G^{H}/ G^{V}$ as a function of $\lambda$ between a cross-shaped crack configuration in the center of the thin layer and a cross-shaped configuration near the fixtures.}\label{Fig4}
\end{figure}

Next, the effect of the degree of confinement of the thin layer on load-parallel crack propagation is evaluated.  
Fig.~\ref{Fig4}(a) presents the results of the energy release rate computations for a nearly incompressible Neo-Hookean material with a cross-shaped crack configuration at the center.  
The results for $G^H/G^V$ are computed for three aspect ratios, $D/H=5$, $D/H=10$, and $D/H=20$, and are plotted as a function of $\lambda$.  
It is observed that for both $D/H=20$ and $D/H=10$, there exists a large range of $\lambda$ values (and correspondingly of $G_c$) for which vertical crack propagation is preferable.  
In contrast, for $D/H=5$, horizontal crack propagation is consistently preferable.  
However, for large values of $G_c$, both cracks may propagate simultaneously as $G^H/G^V \approx 1$.  
The corresponding results for an Ogden hyperelastic material \cite{Ogden72} with a strain stiffening parameter $\alpha=4$ are shown in Fig.~\ref{Fig4}(b).  
The results are qualitatively similar to those for the Neo-Hookean material.
%


For completeness, we also evaluate the case in which a crack initiates near one of the fixtures and develops two crack fronts. The results for $G^H/G^V$ as a function of $\lambda$ for $D/H = 5$ show that vertical crack growth is more favorable in this scenario. This is consistent with the phase-field simulation shown in Fig.~\ref{Fig1}(e) for a disk with $D/H = 5$ made of natural rubber, where nucleation occurs near the fixtures and crack growth initially proceeds in the vertical direction.


\subsection{Growth of the horizontal and vertical cracks}

\begin{figure}[h!]
	\centering
	\includegraphics[width=6.in]{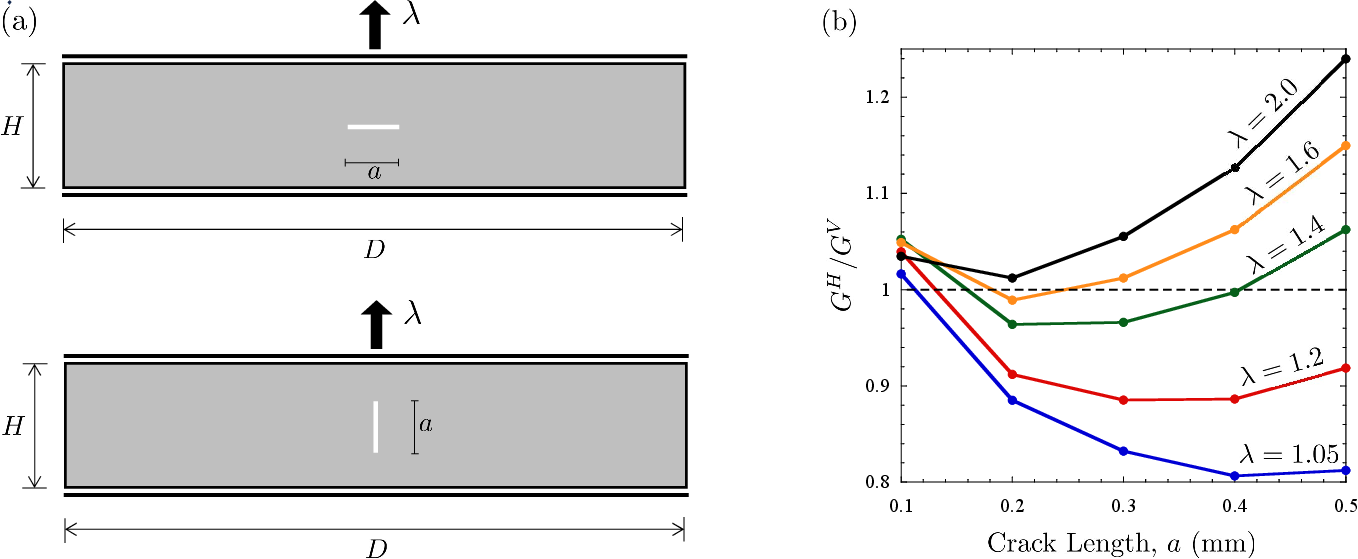}
	\caption{Energy release rate evolution for horizontal and vertical cracks as a function of crack length, $a$, and applied stretch, $\lambda$, for $D/H=10$ and $\nu=0.4995$. (a) Schematic of the geometries, and (b) $G^H/G^V$ as a function of the crack length and applied stretch.}\label{Fig6}
\end{figure}

The above results illustrate crack nucleation from a cross-shaped configuration with crack lengths $a=0.1$ mm in both directions.  We next examine the effect of varying horizontal and vertical crack lengths. For simplicity, we evaluate the energy release rates using two separate geometries: one with a horizontal crack and the other with a vertical crack, as illustrated in the schematic in Fig.~\ref{Fig6}(a). Results shown in Fig.~\ref{Fig6}(b) for $D/H=10$ indicate that for very small and very large cracks, horizontal cracks exhibit a greater tendency to grow---especially when $G_c$ is high and failure occurs at large values of $\lambda$. For intermediate crack lengths and small values of $G_c$, vertical cracks will likely have a faster growth rate. Analysis with simultaneous vertical and horizontal growth can provide further insight.
The analysis could also be extended to study the competition between the nucleation of new cracks and the growth of the existing vertical or horizontal cracks. An example of such an analysis, focusing on the nucleation of a new crack when a \emph{horizontal} pre-crack is present, was described in \cite{ruihuang23cavitation}. 


%

In conclusion, this study has firmly established the relationship between various material and geometrical parameters and the load-parallel crack propagation, which influences the two failure mechanisms observed in thin confined layers. These insights can be applied to control crack propagation in thin films and adhesives, making them stronger and tougher.

\vspace{0.2cm}





\bibliographystyle{elsarticle-num-names}
\bibliography{ref}

\begin{thebibliography}{17}
\expandafter\ifx\csname natexlab\endcsname\relax\def\natexlab#1{#1}\fi
\providecommand{\url}[1]{\texttt{#1}}
\providecommand{\href}[2]{#2}
\providecommand{\path}[1]{#1}
\providecommand{\DOIprefix}{doi:}
\providecommand{\ArXivprefix}{arXiv:}
\providecommand{\URLprefix}{URL: }
\providecommand{\Pubmedprefix}{pmid:}
\providecommand{\doi}[1]{\href{http://dx.doi.org/#1}{\path{#1}}}
\providecommand{\Pubmed}[1]{\href{pmid:#1}{\path{#1}}}
\providecommand{\bibinfo}[2]{#2}
\ifx\xfnm\relax \def\xfnm[#1]{\unskip,\space#1}\fi
\bibitem[{Gent and Lindley(1959)}]{GentLindley1959}
\bibinfo{author}{A.~Gent}, \bibinfo{author}{P.~Lindley},
\newblock \bibinfo{title}{Internal rupture of bonded rubber cylinders in tension},
\newblock \bibinfo{journal}{Proceedings of the Royal Society of London. Series A. Mathematical and Physical Sciences} \bibinfo{volume}{249} (\bibinfo{year}{1959}) \bibinfo{pages}{195--205}.
\bibitem[{Lakrout et~al.(1999)Lakrout, Sergot, and Creton}]{creton1999direct}
\bibinfo{author}{H.~Lakrout}, \bibinfo{author}{P.~Sergot}, \bibinfo{author}{C.~Creton},
\newblock \bibinfo{title}{Direct observation of cavitation and fibrillation in a probe tack experiment on model acrylic pressure-sensitive-adhesives},
\newblock \bibinfo{journal}{The Journal of Adhesion} \bibinfo{volume}{69} (\bibinfo{year}{1999}) \bibinfo{pages}{307--359}.
\bibitem[{Creton and Lakrout(2000)}]{creton2000flatprobe}
\bibinfo{author}{C.~Creton}, \bibinfo{author}{H.~Lakrout},
\newblock \bibinfo{title}{Micromechanics of flat-probe adhesion tests of soft viscoelastic polymer films},
\newblock \bibinfo{journal}{Journal of Polymer Science Part B: Polymer Physics} \bibinfo{volume}{38} (\bibinfo{year}{2000}) \bibinfo{pages}{965--979}.
\bibitem[{Crosby et~al.(2000)Crosby, Shull, Lakrout, and Creton}]{crosbyshullcreton2000}
\bibinfo{author}{A.~J. Crosby}, \bibinfo{author}{K.~R. Shull}, \bibinfo{author}{H.~Lakrout}, \bibinfo{author}{C.~Creton},
\newblock \bibinfo{title}{Deformation and failure modes of adhesively bonded elastic layers},
\newblock \bibinfo{journal}{Journal of Applied Physics} \bibinfo{volume}{88} (\bibinfo{year}{2000}) \bibinfo{pages}{2956--2966}.
\bibitem[{Griffith(1921)}]{Griffith21}
\bibinfo{author}{A.~A. Griffith},
\newblock \bibinfo{title}{Vi. the phenomena of rupture and flow in solids},
\newblock \bibinfo{journal}{Philosophical transactions of the royal society of london. Series A, containing papers of a mathematical or physical character} \bibinfo{volume}{221} (\bibinfo{year}{1921}) \bibinfo{pages}{163--198}.
\bibitem[{Kamarei et~al.(2024)Kamarei, Kumar, and Lopez-Pamies}]{KKLP24}
\bibinfo{author}{F.~Kamarei}, \bibinfo{author}{A.~Kumar}, \bibinfo{author}{O.~Lopez-Pamies},
\newblock \bibinfo{title}{The poker-chip experiments of synthetic elastomers},
\newblock \bibinfo{journal}{arXiv preprint arXiv:2402.06785}  (\bibinfo{year}{2024}).
\bibitem[{Guo and Ravi-Chandar(2023)}]{guo-ravichandar23}
\bibinfo{author}{J.~Guo}, \bibinfo{author}{K.~Ravi-Chandar},
\newblock \bibinfo{title}{On crack nucleation and propagation in elastomers: I. in situ optical and x-ray experimental observations},
\newblock \bibinfo{journal}{International Journal of Fracture} \bibinfo{volume}{243} (\bibinfo{year}{2023}) \bibinfo{pages}{1--29}.
\bibitem[{Kumar and Lopez-Pamies(2021)}]{KLP21}
\bibinfo{author}{A.~Kumar}, \bibinfo{author}{O.~Lopez-Pamies},
\newblock \bibinfo{title}{The poker-chip experiments of {G}ent and {L}indley (1959) explained},
\newblock \bibinfo{journal}{Journal of the Mechanics and Physics of Solids} \bibinfo{volume}{150} (\bibinfo{year}{2021}) \bibinfo{pages}{104359}.
\bibitem[{Lef{\`e}vre et~al.(2015)Lef{\`e}vre, Ravi-Chandar, and Lopez-Pamies}]{lefevre2015cavitation}
\bibinfo{author}{V.~Lef{\`e}vre}, \bibinfo{author}{K.~Ravi-Chandar}, \bibinfo{author}{O.~Lopez-Pamies},
\newblock \bibinfo{title}{Cavitation in rubber: an elastic instability or a fracture phenomenon?},
\newblock \bibinfo{journal}{International Journal of Fracture} \bibinfo{volume}{192} (\bibinfo{year}{2015}) \bibinfo{pages}{1--23}.
\bibitem[{Poulain et~al.(2017)Poulain, Lefevre, Lopez-Pamies, and Ravi-Chandar}]{Poulain17}
\bibinfo{author}{X.~Poulain}, \bibinfo{author}{V.~Lefevre}, \bibinfo{author}{O.~Lopez-Pamies}, \bibinfo{author}{K.~Ravi-Chandar},
\newblock \bibinfo{title}{Damage in elastomers: nucleation and growth of cavities, micro-cracks, and macro-cracks},
\newblock \bibinfo{journal}{International Journal of Fracture} \bibinfo{volume}{205} (\bibinfo{year}{2017}) \bibinfo{pages}{1--21}.
\bibitem[{Kumar et~al.(2018)Kumar, Francfort, and Lopez-Pamies}]{KFLP18}
\bibinfo{author}{A.~Kumar}, \bibinfo{author}{G.~A. Francfort}, \bibinfo{author}{O.~Lopez-Pamies},
\newblock \bibinfo{title}{Fracture and healing of elastomers: A phase-transition theory and numerical implementation},
\newblock \bibinfo{journal}{Journal of the Mechanics and Physics of Solids} \bibinfo{volume}{112} (\bibinfo{year}{2018}) \bibinfo{pages}{523--551}.
\bibitem[{Breedlove et~al.(2024)Breedlove, Chen, Lindeman, and Lopez-Pamies}]{breedlove2024}
\bibinfo{author}{E.~Breedlove}, \bibinfo{author}{C.~Chen}, \bibinfo{author}{D.~Lindeman}, \bibinfo{author}{O.~Lopez-Pamies},
\newblock \bibinfo{title}{Cavitation in elastomers: A review of the evidence against elasticity},
\newblock \bibinfo{journal}{Journal of the Mechanics and Physics of Solids}  (\bibinfo{year}{2024}) \bibinfo{pages}{105678}.
\bibitem[{Kumar et~al.(2020)Kumar, Bourdin, Francfort, and Lopez-Pamies}]{KBFLP20}
\bibinfo{author}{A.~Kumar}, \bibinfo{author}{B.~Bourdin}, \bibinfo{author}{G.~A. Francfort}, \bibinfo{author}{O.~Lopez-Pamies},
\newblock \bibinfo{title}{Revisiting nucleation in the phase-field approach to brittle fracture},
\newblock \bibinfo{journal}{Journal of the Mechanics and Physics of Solids} \bibinfo{volume}{142} (\bibinfo{year}{2020}) \bibinfo{pages}{104027}.
\bibitem[{Hao et~al.(2023)Hao, Suo, and Huang}]{ruihuang23cavitation}
\bibinfo{author}{S.~Hao}, \bibinfo{author}{Z.~Suo}, \bibinfo{author}{R.~Huang},
\newblock \bibinfo{title}{Why does an elastomer layer confined between two rigid blocks grow numerous cavities?},
\newblock \bibinfo{journal}{Journal of the Mechanics and Physics of Solids} \bibinfo{volume}{173} (\bibinfo{year}{2023}) \bibinfo{pages}{105223}.
\bibitem[{Francfort and Marigo(1998)}]{Francfort98}
\bibinfo{author}{G.~A. Francfort}, \bibinfo{author}{J.-J. Marigo},
\newblock \bibinfo{title}{Revisiting brittle fracture as an energy minimization problem},
\newblock \bibinfo{journal}{Journal of the Mechanics and Physics of Solids} \bibinfo{volume}{46} (\bibinfo{year}{1998}) \bibinfo{pages}{1319--1342}.
\bibitem[{Ogden(1972)}]{Ogden72}
\bibinfo{author}{R.~W. Ogden},
\newblock \bibinfo{title}{Large deformation isotropic elasticity--on the correlation of theory and experiment for incompressible rubberlike solids},
\newblock \bibinfo{journal}{Proceedings of the Royal Society of London. A. Mathematical and Physical Sciences} \bibinfo{volume}{326} (\bibinfo{year}{1972}) \bibinfo{pages}{565--584}.
\bibitem[{Rice(1968)}]{rice1968path}
\bibinfo{author}{J.~Rice},
\newblock \bibinfo{title}{A path independent integral and the approximate analysis of strain concentration by notches and cracks},
\newblock \bibinfo{journal}{Journal of Applied Mechanics} \bibinfo{volume}{35} (\bibinfo{year}{1968}) \bibinfo{pages}{379}.

\end{thebibliography}

\end{document}